\documentclass[useAMS,usenatbib]{mn2e}
\usepackage{url}
\usepackage{graphicx}
\usepackage{verbatim}
\usepackage[usenames]{color}
\usepackage[flushleft]{threeparttable}
\DeclareGraphicsExtensions{.pdf,.png,.jpg,.mps,.eps,.ps}
\usepackage{amsmath}
\usepackage{natbib}
\usepackage{color}
\usepackage{graphicx}
\usepackage{lscape}
\usepackage{gensymb}
\usepackage{amssymb}
\usepackage{psfrag}

\newcommand\nicer{{\it NICER}}
\newcommand\nustar{{\it NuSTAR}}
\newcommand\swift{{\it SWIFT}}

\newcommand\rxte{{\it RXTE}}

\newcommand\inte{{\it INTEGRAL}}

\newcommand\kev{{\rm~keV}}

\newcommand\kms{\ifmmode {\rm~km\ s}^{-1} \else ~km s$^{-1}$\fi}
\newcommand\Hunit{\ifmmode {\rm~km\ s}^{-1}\ {\rm Mpc}^{-1}
        \else ~km s$^{-1}$ Mpc$^{-1}$\fi}
\newcommand\ctssec{\ifmmode {\rm~count\ s}^{-1} \else ~count s$^{-1}$\fi}
\newcommand\ergsec{\ifmmode {\rm~erg\ s}^{-1} \else
        ~erg s$^{-1}$\fi}
\newcommand\funit{\ifmmode {\rm~erg\ s}^{-1}\;{\rm cm}^{-2} \else
        ~ergs s$^{-1}$ cm$^{-2}$\fi}
\newcommand\phflux{\ifmmode {\rm~photon\ s}^{-1}\;{\rm cm}^{-2}
        \else   ~photon s$^{-1}$ cm$^{-2}$\fi}
\newcommand\efluxA{\ifmmode {\rm~erg\ s}^{-1}\;{\rm cm}^{-2}\;{\rm
        \AA}^{-1} \else ~erg s$^{-1}$ cm$^{-2}$ \AA$^{-1}$\fi}
\newcommand\efluxHz{\ifmmode {\rm~erg\ s}^{-1}\;{\rm cm}^{-2}\;{\rm
        Hz}^{-1} \else ~erg s$^{-1}$ cm$^{-2}$ Hz$^{-1}$\fi}
\newcommand\cc{\ifmmode {\rm~cm}^{-3} \else cm$^{-3}$\fi}
\newcommand\FWHM{\ifmmode {\rm~FWHM} \else ${\rm~FWHM}$\fi}
\newcommand\Msun{\ifmmode M_{\odot} \else $M_{\odot}$\fi}
\newcommand\Lsun{\ifmmode L_{\odot} \else $L_{\odot}$\fi}

\newcommand\hbeta{\ifmmode {\rm H}\beta \else H$\beta$\fi}
\newcommand\Kalpha{\ifmmode {\rm K}\alpha \else K$\alpha$\fi}
\newcommand\nh{\ifmmode N_{\rm H} \else N$_{\rm H}$\fi}

\usepackage{graphicx}
\usepackage{url}
\usepackage{verbatim}
\usepackage{color}
\usepackage[colorlinks,citecolor=blue,urlcolor=blue,bookmarks=false,hypertexnames=true,pagebackref=true]{hyperref} 

\title[Disc reflection from IGR~J17498-2921]{Relativistic X-ray reflection from the accreting millisecond X-ray pulsar IGR~J17498-2921}
\author[Mondal et al.]{\parbox[]{6.5in}{Mahasweta Bhattacharya$^{1}$, Aditya S. Mondal$^{1}\thanks{E-mail: adityas.mondal@visva-bharati.ac.in}$, Mayukh Pahari$^{2}$, Biplab Raychaudhuri$^{1}$, Rohit Ghosh$^{1}$, Gulab C. Dewangan$^{3}$   \\
\small
$^{1}$Department of physics, Visva-Bharati, Santiniketan, West Bengal, 731235, India \\
$^{2}$Department of Physics, Indian Institute of Technology Hyderabad, Hyderabad, Kandi, 502285 Sangareddy, India \\
$^{3}$Inter-University Centre for  Astronomy \& Astrophysics (IUCAA), Pune, 411007, India \\
}}

\date{\today}
\begin{document}
\maketitle
\begin{abstract}
The accreting millisecond X-ray pulsar IGR~J17498-2921 went into X-ray outburst on April 13-15, 2023, for the first time since its discovery on August 11, 2011. Here, we report on the first follow-up \nustar{} observation of the source, performed on April 23, 2023, around ten days after the peak of the outburst. The \nustar{} spectrum of the persistent emission ($3-60$ \kev{} band) is well described by an absorbed blackbody with a temperature of $kT_{bb}=1.61\pm 0.04$\kev{}, most likely arising from the NS surface and a Comptonization component with power-law index $\Gamma=1.79\pm0.02$, arising from a hot corona at $kT_{e}=16\pm 2$ keV. The X-ray spectrum of the source shows robust reflection features which have not been observed before. We use a couple of self-consistent reflection models, {\tt relxill} and {\tt relxillCp}, to fit the reflection features. We find an upper limit to the inner disc radius of $ 6\: R_{ISCO}$ and $ 9\: R_{ISCO}$ from {\tt relxill} and {\tt relxillCp} model, respectively. The inclination of the system is estimated to be $\simeq 40\degr$ from both reflection models. Assuming magnetic truncation of the accretion disc, the upper limit of magnetic field strength at the pole of the NS is found to be $B\lesssim 1.8\times 10^{8}$ G. Furthermore, the \nustar{} observation revealed two type I X-ray bursts and the burst spectroscopy confirms the thermonuclear nature of the burst. The blackbody temperature reaches nearly $2.2$ keV at the peak of the burst.
 
\end{abstract}

\section{Introduction}
An accreting millisecond pulsar (AMSP) is a Neutron star (NS) accreting mass from a low-mass ($\lesssim 1 \Msun$) donor/companion star and showing a coherent signal with a period of few ms \citep{2018ASSL..457..149C, 2022ASSL..465...87D}. They are a high observational priority due to their rarity. To date, only 27 AMSPs have been discovered since 1998. SAX~J1808.4-3658 is the first discovered AMSP \citep{1998ApJ...507L..63W}, the recent ones being MAXI~J1816-195 \citep{2022ATel15425....1B} and MAXI~J1957+032 \citep{2022ATel15444....1N}. Very recently \citet{2024ATel16464....1M} reported the discovery of a AMSP SRGA~J144459.2-604207 on February 21, 2024, by the ART-XC telescope. NS low-mass X-ray binaries (LMXBs) emit X-rays due to the accretion of matter from the companion star onto the NS through Roche-lobe overflow. The pulsation is caused by the NS magnetic field, which is strong enough to funnel the accreting matter to the magnetic poles effectively. NS LMXBs are classified into persistent and transient sources based on long-term X-ray variabilities. Persistent NS LMXBs accreate matter continuously and may have an X-ray luminosity of $L_{X}\gtrsim 10^{36} \ergsec{}$ (\citealt{2019ApJ...873...99L, 2017ApJ...836..140L}). On the contrary, transient LMXBs do not accrete matter continuously and undergo recurrent bright ($L_{X}\gtrsim 10^{36} \ergsec{}$) outbursts lasting from days to weeks and then return to prolonged intervals of X-ray quiescence ($L_{X}\lesssim 10^{34} \ergsec{}$) lasting from months to years \citep{2010A&A...524A..69D, 2020arXiv201009005D}. Most of the AMSPs are transient X-ray sources with recurrence times between two and more than ten years, and their outbursts usually last from a week to a few months \citep{Patruno_2020}. Almost all AMSPs have short, i.e., $\sim$hours or minutes, orbital periods and are therefore characterized by compact orbits. Most of these systems are bursters, as they have displayed a type-I thermonuclear X-ray burst at least once (\citealt{2021ASSL..461..209G, 2010MNRAS.407.2575P, 2019MNRAS.483..767D, 2022ApJ...935L..32B, 2024ApJ...968L...7N}). The bursts from AMSPs exhibit burst oscillations at the typical spin frequency \citep{Patruno_2020}.  \\

From a spectral point of view, the emission from these systems in outburst is usually dominated by the Comptonization spectrum from a hot corona with electron temperatures usually of tens of keV \citep{2022ASSL..465...87D}. AMSPs in outburst are, therefore, (almost) always in a hard spectral state. The accretion flow is expected to stop far from the NS surface in this state. However, two AMSPs, SAX~J1748.9-2021 and SAX~J1808.4-3658, have been observed to transition into the soft states \citep{2016MNRAS.457.2988P, 2019MNRAS.483..767D}. Besides an energetically dominating Comptonized component, one or two soft components with relatively low temperatures are often detected depending on the statistics and spectral resolution of the detectors. These soft components are interpreted as the emission arising from the accretion disc and the NS surface/boundary layer. An additional spectral component arises when the Comptonization spectrum emitted by the corona illuminates the disc and is reprocessed by it. This component is known as the reflection spectrum, a prevalent ingredient in spectra of X-ray binaries \citep{1989MNRAS.238..729F}. Disk reflection components are mainly characterized by a Fe-K emission line at $\simeq 6.4–6.97$ keV and a Compton hump at $\simeq 20–40$ keV \citep{1989MNRAS.238..729F}. X-ray reflection spectroscopy provides a powerful diagnostic tool for investigating the dynamics and geometry of the accretion disk. Detecting and modeling disk reflection features allows for a measure of the inner radial extent of the accretion disk, the ionization of the plasma in the disc, and the inclination of the system. Reflection features have been observed in some AMSPs for which data with high-to-moderate energy resolution were available \citep{2013MNRAS.429.3411P, 2017MNRAS.471..463S, 2019MNRAS.483..767D, 2022MNRAS.515.3838M} but not all \citep{2005A&A...436..647F, 2018A&A...610L...2S, 2018A&A...616L..17S}.\\

IGR~J17498-2921 is known as an AMSP, discovered with \inte{} on August 11, 2011 \citep{2011ATel.3551....1G}. The source position was determined by follow-up Swift \citep{2011ATel.3558....1B} and Chandra \citep{2011ATel.3606....1C} observations. The source is located at $\alpha=17^{h}49^{m}55^{s}.35$, $\delta=-29^{\degr}19^{'}19.6^{''}$, with an associated uncertainty of $0.6^{''}$ at the $90\%$ confidence level. Pulsation at a frequency of $\sim 401$ Hz was observed from this source \citep{2011ATel.3556....1P}. The source resides in a 3.8 hr binary \citep{2011ATel.3561....1M}, orbiting a low-mass companion donor with mass $\sim 0.2 \Msun$ \citep{2024arXiv240316471G}. Thermonuclear X-ray bursts were detected by \citet{2011ATel.3560....1F} during \inte{} observations of this source. \citet{2011ATel.3568....1L} reported burst oscillations from this source at a frequency consistent with the spin frequency. Burst oscillation from this source was further confirmed by \citet{2011ATel.3643....1C} using \rxte{}/PCA data with a much higher significance. Bursts from this source also show the characteristics of the photospheric radius expansion episode, which gives a distance estimate of $\sim 7.6$ kpc \citep{2011ATel.3568....1L}. After a 37-day-long outburst, the source returned to quiescence on September 19, 2011 \citep{2011ATel.3661....1L}. The source was observed at a luminosity of $\sim 2\times 10 ^{32}$ \ergsec{} during quiescence \citep{2011ATel.3559....1J}. From the current \inte{} observation performed on April 13-15, 2023, the source was again found in a bright X-ray state for the first time since its discovery in 2011 \citep{2023ATel15996....1G}. \nicer{} observation also confirms the new outburst from this source \citep{2023ATel15998....1S}. \nicer{} follow-up observations starting on April 19, 2023, detected the source at a constant count rate of $\sim 80$ ct/s in the $0.5-10$ keV band \citep{2023ATel15998....1S}. Later, \nustar{} observed the source on April 23, 2023, around ten days after the peak of the outburst. We used the \nustar{} observation for our analysis.\\

\citet{2011A&A...535L...4P} analyzed the spectral properties of IGR~J17498-2921 based on the observations performed by the \swift{} and the \rxte{}/PCA between August 12 and September 22, 2011. During most of the outburst, the spectra are modeled by a power-law with an index $\Gamma \approx 1.7-2$, while values of $\approx 3$ are observed as the source fades into quiescence. Later, \citet{2012A&A...545A..26F} studied the broad-band spectrum of the persistent emission in the $0.6 – 300$ keV energy band using simultaneous \inte{}, \rxte{}, and \swift{} data obtained in August–September 2011. They also discussed the properties of the largest set of X-ray bursts from this source discovered by these satellites. They found that the broad-band average spectrum is well-described by thermal Comptonization with an electron temperature of $kT_{e}\sim 50$ keV, soft seed photons of $kT_{bb} \sim 1$ keV, and Thomson optical depth $\tau_{T} \sim 1$ in slab geometry. The \inte{}, \rxte{}, and \swift{} data also reveal the X-ray pulsation at a period of $2.5$ ms up to $\sim 65$ keV \citep{2011A&A...535L...4P}. \citet{2023ATel15998....1S} reported on the recent \nicer{} observation of this source performed on April 19, 2023 and they found that the $1-10$ keV energy spectrum is well described using an absorbed blackbody and a Comptonization continuum ({\tt nthcomp}). They also found a blackbody temperature of $\sim 0.30$ keV, and a Comptonization power-law index of $\sim 1.76$. They estimated the $0.5-10$ keV unabsorbed flux at $\sim 1\times 10^{-9}$ erg~s$^{-1}$ cm$^{-2}$, which matches the 2011 peak flux. \\

In this work, we represent the spectral analysis of the persistent emission of this source using the available \nustar{} observation performed on April 23, 2023. We also investigate the properties of the type-I X-ray bursts from this source. Due to unprecedented sensitivity above 10 keV, it is possible to constrain the accretion geometry of the system by modeling the reflection spectrum (Fabian et al.1989) using \nustar{}. Moreover, fitting with self-consistent reflection models will give important information on the position of the magnetospheric radius and the neutron star magnetic field. We have organized the paper as follows: we provide an overview of the observation and data reduction in Section 2. We present the source light curves in Section 3. We provide the details of the spectral analysis, both persistent and burst emission, in Section 4. Finally, we discuss the results obtained from the analysis in Section 5.

\begin{figure*}
\centering
\includegraphics[scale=0.50, angle=0]{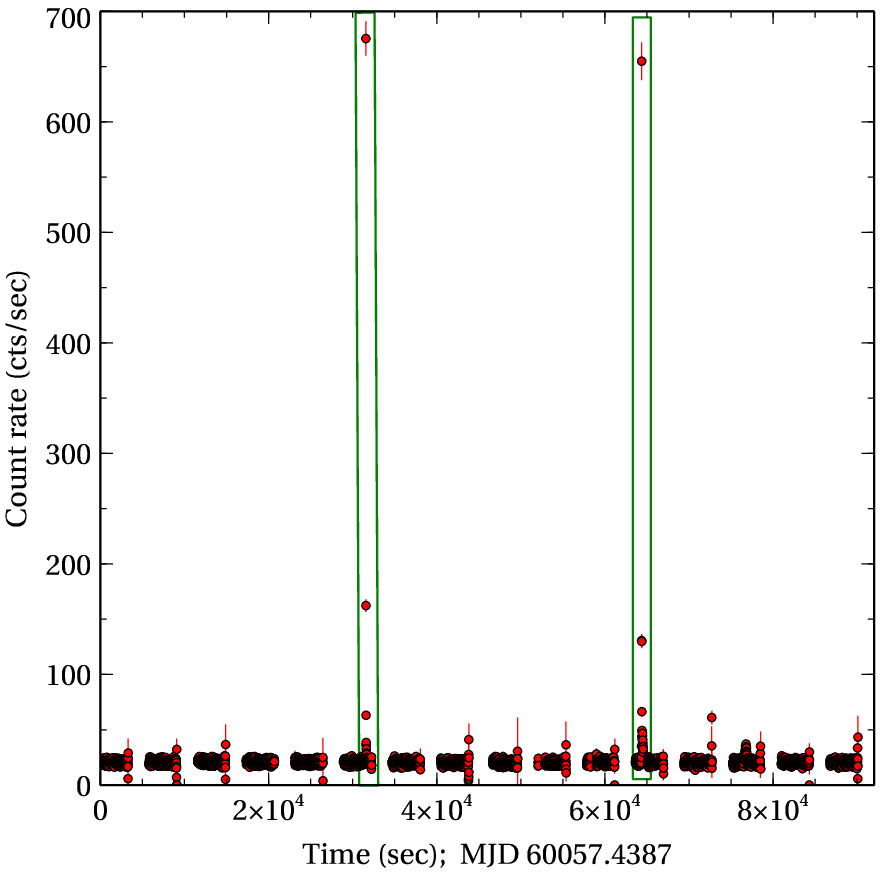}\hspace{2cm}
\includegraphics[scale=0.50, angle=0]{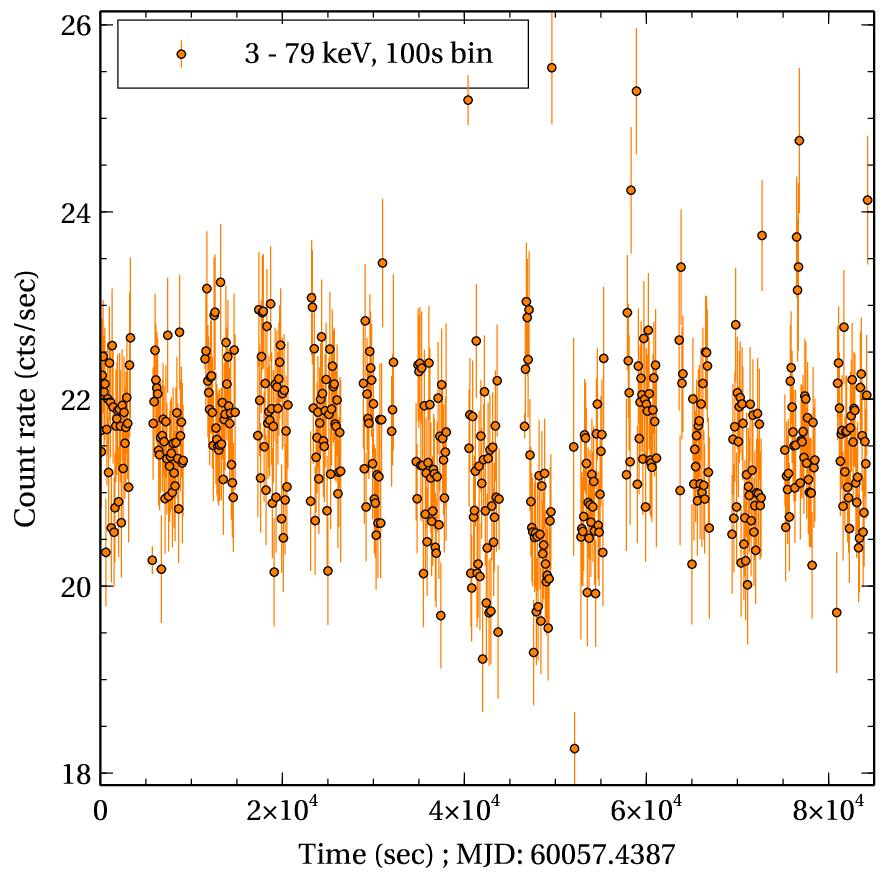}
\caption{Left panel: $3-79\kev{}$ \nustar{}/FPMA light curve of the source with a binning of 10s. The source shows the presence of two thermonuclear X-ray burst. Right panel: Same light curve with a binning of 100s after removing burst intervals.} 
\label{Fig1}
\end{figure*}

\section{observation and data reduction}
\textit{Nuclear Spectroscopic Telescope ARray} (\nustar{}; \citealt{2013ApJ...770..103H}) observed the source IGR~J17498-2921 only once on April 23, 2023 for a total exposure of $\sim 45$ ks. We have used this observation (obsID: 90901317002) for our analysis.  \\

The \nustar{} data were collected in the $3-79$\kev{} energy band using two identical co-aligned telescopes equipped with the focal plane modules FPMA and FPMB. We reduced the data using the standard data analysis software {\tt NUSTARDAS v2.1.2} task included in {\tt HEASOFT v6.33.2}. We used the latest \nustar{} {\tt CALDB} version available ($v20240206$) at the time of the analysis. We used the task {\tt nupipeline v0.4.9} to generate the calibrated and screened event files. 
The source events have been extracted from a circular region centered on the source position, with a radius of $120''$ for both detectors FPMA and FPMB. The background events also have been extracted from a circle of the same radius from the area of the same chip but in a position far from the source for both instruments. 
The filtered event files, the background subtracted light curves, and the spectra for both detectors are extracted using the tool {\tt nuproducts}. Corresponding response files are also created as output of {\tt nuproducts}, which ensures that all the instrumental effects, including loss of exposure due to dead-time, are correctly accounted for. We grouped the FPMA and FPMB spectral data with a minimum of $100$ counts per bin which allows the use of $\chi^{2}$ statistics. Finally, we fitted spectra from detectors FPMA and FPMB simultaneously, leaving a floating cross-normalization constant.

\section{Light Curve}
In the left panel of Figure~\ref{Fig1}, we show the \nustar{}/FPMA light curve of the source in the $3-79$ \kev{} energy band during the decay of the 2023 outburst. The average count rate during the \nustar{} observation was $\sim 20-24$ counts s$^{-1}$ (see right panel of Figure~\ref{Fig1}). Two type-I thermonuclear X-ray bursts were present at about $\sim 31$th ks and $\sim 64$th ks during the \nustar{} observation. Those are marked with the green vertical boxes. This is the first detection of X-ray bursts in the 2023 outburst cycle as no bursts have been reported by the prevoius \inte{} and \nicer{} observations. The peak of the burst seems to reach approximately $\sim 680$ counts s$^{-1}$, about a factor of $\sim 28$, the level of the persistent emission. Both type-I bursts lasted about $\sim 20$ s. We have extracted the light curve and spectrum of the bursts and investigated the properties in details. For the persistent emission spectrum, we eliminated a time interval of around $1000$ s surrounding the burst, starting from $250$ s before the rise of the burst. 

\begin{figure*}
\includegraphics[scale=0.45, angle=-90]{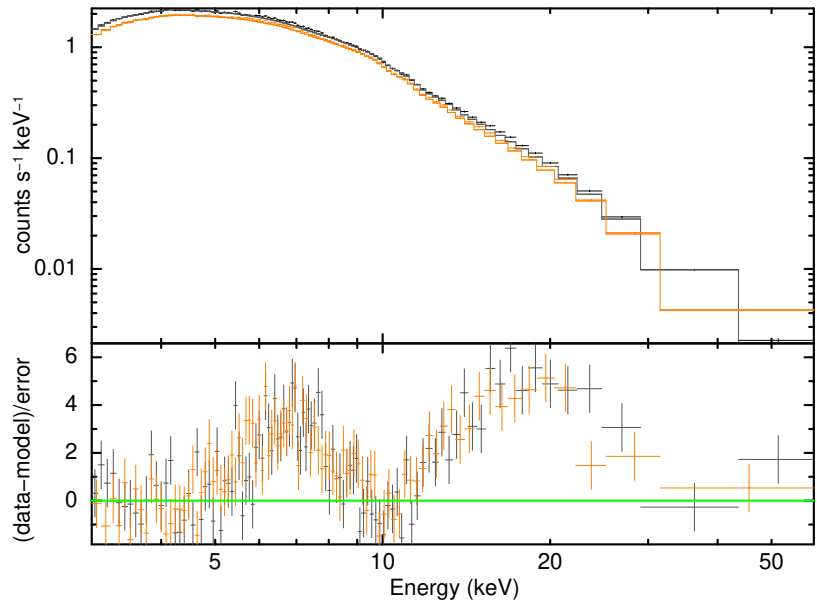}
\caption{The source spectrum in the energy band $3-60\kev{}$ obtained from the \nustar{} FPMA and FPMB is presented here. The continuum emission is fitted with the absorbed {\tt bbody} and the  {\tt nthcomp} model. Fit ratio associated with this continuum model is shown in the bottom panel of this plot. The presence of disc reflection is evident in the spectrum.} 
\label{Fig3}
\end{figure*}

\section{spectral analysis}
We used the X-ray spectral package {\tt XSPEC} $v12.14.0h$ \citep{1996ASPC..101...17A} to model the \nustar{} FPMA and FPMB spectra simultaneously between $3$ to $60$\kev{} energy band, while above $60$\kev{} spectrum is dominated by the background. A constant multiplication factor {\tt constant} was included in the modeling to account for cross-calibration of the two instruments, FPMA and FPMB. The value of {\tt constant} for FPMA was fixed to $1$, allowing it to vary for the FPMB. We used the {\tt TBabs} model to account for interstellar absorption along the line of sight with the {\tt wilm} abundances \citep{2000ApJ...542..914W} and the {\tt vern} \citep{1996ApJ...465..487V} photoelectric cross-section. Unless otherwise stated, spectral uncertainties are quoted at the $1 \sigma$ confidence level.

\subsection{Continuum modeling}
Different combinations of model components and spectral parameters are required to describe the continuum emission from different spectral states from the NS LMXBs. To probe the spectral shape of this source correctly, we first tried to fit the continuum emission (\nustar{} spectrum $3-60$\kev{}) with a model consisting of an absorbed cutoff power-law ({\tt cutoffpl}) and a single-temperature blackbody component ({\tt bbody}). These models describe the emission from the presence of the corona and surface of the NS and/or boundary layer region, respectively. This standard model combination has previously been used to fit the broad-band continuum emission in AMSPs \citep{2022ASSL..465...87D}. This two-component model describes the continuum emission reasonably well. The poor quality of the fit ($\chi^2/dof=1561/1273$) is mainly due to the presence of strong reflection features in the $5-9$ \kev{} and $12-30$ \kev{} energy range. Because of the lack of sensitivity of \nustar{} at low energies, we had to fix the photoelectric equivalent hydrogen column density at $0.81 \times 10^{22}$ cm$^{-2}$, following \citet{1990ARA&A..28..215D}. We found that the spectrum is dominated by a hard Comptonization component from the corona with a photon index of $\Gamma=1.70\pm 0.02$ and the cutoff energy $E_{cut}= 55\pm 4$\kev{}. The cutoff power-law component accounts for $\simeq 88\%$ of the total $3 - 79$ \kev{} unabsorbed flux. The inferred blackbody temperature $kT_{bb}= 1.55\pm 0.03\kev{}$ suggests that this continuum may originate from the NS surface and/or boundary layer region.\\

A much better description of continuum emission due to thermal Comptonization is given by the model {\tt nthcomp} \citep{1996MNRAS.283..193Z}. It attempts to simulate the upscattering of photons through the corona from a seed spectrum parameterized either by the inner temperature of the accretion disc or the NS surface/boundary layer. Both shapes can be selected via the {\it input type}, parameterized by a seed photon temperature. The high energy cutoff is parameterized by the electron temperature of the medium. We then replaced the cutoff power-law model with the physically-motivated Comptonization model {\tt nthcomp}, setting the photon seed input to a disc blackbody. This model also describes the continuum emission very well with $\chi^2/dof=1495/1272$. We estimated the electron temperature at $kT_{e}= 16\pm 2\kev{}$ with photon index $\Gamma= 1.79\pm 0.02$. The seed photon temperature is found to be low $kT_{seed}\lesssim 0.31\kev{}$, indicating the difficulties in detecting the additional soft disc blackbody component separately arising from the accretion disc. The electron temperature of the corona, $kT_{e}$, is roughly comparable to the observed cutoff energy ($\sim 55$\kev{}) of the spectrum. The optical depth of the corona is found to be $\tau= 4.0\pm 0.2$ for the {\tt nthcomp} model, suggesting an optically thin corona. Therefore, seed photons of temperature lower than $\approx 0.31$\kev{} are comptonized in a corona with an electron temperature of $\sim 16$\kev{} and an optical depth of $\sim 4.0$, corresponding to a photon index of the Comptonized spectrum of $\Gamma\sim 1.8$. \\

Our continuum description with different combinations of the model leaves large positive residuals around $5-9$\kev{} and $12-30$\kev{}, as shown in Figure~\ref{Fig3}. It indicates the presence of a broad iron emission feature and Compton reflection hump, respectively. These features are typically attributed to disc reflection. This motivates the inclusion of a relativistic reflection model in our spectral fits.\\

\begin{figure*}
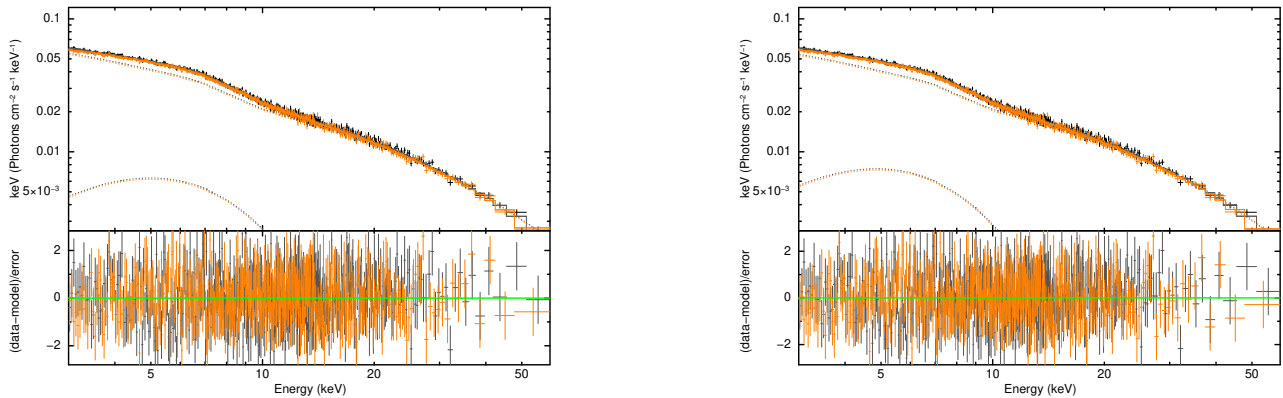

\includegraphics[scale=0.30, angle=-90]{fig4.ps}\hspace{2cm}
\includegraphics[scale=0.30, angle=-90]{fig5.ps}
\caption{ Left: The unfolded spectral data, the best-fit model {\tt const*TBabs*(bbody+RELXILL)}. The lower panel of this plot shows the ratio of the data to the model in units of $\sigma$. Right: The unfolded spectral data, the best-fit model {\tt const*TBabs*(bbody+RELXILLCP)} } 
\label{Fig4}
\end{figure*}

\subsection{Self-consistent reflection fitting}
\subsubsection{Reflection fits with {\tt relxill}}
When hard X-rays irradiate the accretion disc, they produce a reflection spectrum that includes fluorescence lines, recombination, and other emission \citep{1989MNRAS.238..729F}. The shape of the reflection spectrum depends on the properties of the flux incident on the accretion disk. In most X-ray sources, the illuminating continuum input is generally a hard power-law spectrum. However, in NS systems, the emission from the NS surface/boundary layer may be significant and contribute to the reflection \citep{2008ApJ...674..415C, 2019ApJ...873...99L, 2017ApJ...836..140L}. In the case of IGR~J17498-2921, the X-ray spectrum is dominated by a power-law component. We choose the reflection model {\tt relxill} to reproduce both Comptonization and reflection emission correctly. It has a cut-off power-law with photon index $\Gamma$ and cut-off energy $E_{cut}$ as the illuminating photon distribution. The model combines the reflection grid {\tt xillver} \citep{2013ApJ...768..146G} with the convolution kernel {\tt relconv} to include relativistic effects on the shape of the reflection spectrum \citep{2010MNRAS.409.1534D}. \\

The model parameters of {\tt relxill} are as follows: the inner and outer disc emissivity indices $q_{1}$ and $q_{2}$, respectively, the break radius $R_{break}$ which sets the location in the disc where the emissivity index changes from $q_{1}$ to $q_{2}$, the inner and outer radii of the disc $R_{in}$ and $R_{out}$ in units of the ISCO ($R_{ISCO}$), respectively, the inclination angle $i$ of the system wrt our line of sight, the dimensionless spin parameter $a$, the photon index of the input cutoff power-law $\Gamma$, the ionization parameter log$\xi$, the iron abundance of the system $A_{Fe}$, the cutoff energy $E_{cut}$, the reflection fraction $r_{refl}$, and the normalization of the model. During our fitting, we used a single emissivity profile (hence $R_{break}$ is obsolete). We fixed the emissivity index at $q_{1}=q_{2}=3$ (a value commonly found in X-ray binaries). The redshift of the source ($z$) is fixed at $0$ as the source is Galactic one. We fixed $a=0.19$ as the source has a spin frequency of $401$ Hz (considering $a\simeq 0.47/P_{ms}$ \citep{2000ApJ...531..447B} where $P_{ms}$ is the spin period in ms). The outer disc radius is fixed at $R_{out}=1000\;R_{g}$ (where $R_{g}=GM/c^2)$. The parameters $R_{in}$, $i$, $A_{Fe}$, log$\xi$, $\Gamma$, $E_{cut}$, $r_{refl}$ were left free to vary. The addition of the model {\tt relxill} improved the fit significantly to $\chi^2/dof=1244/1268$ ($\Delta\chi^2=-317$ for the $5$ additional parameters). The corresponding spectrum for the model {\tt constant*TBabs*(bbody+relxill)} and the residuals are shown in the left panel of Figure~\ref{Fig4}. The best-fit parameters are reported in Table~\ref{parameters1}\\

\subsubsection{Reflection fits with {\tt relxillCp}}
There is a flavor of relxill with a Comptonization illuminating continuum known as {\tt relxillCp}. The primary emission for this model is given by the Comptonized model {\tt nthcomp}. However, model {\tt relxillCp} assumes the seed photons arise from the accretion disk with a fixed temperature of $0.05$ keV \citep{2014MNRAS.444L.100D}. Since we find a low seed photon temperature lower than $\sim 0.3 $ \kev{} in our continuum fit with {\tt nthcomp}, we attempt to use {\tt relxillCp} in our spectral fit. The model {\tt relxillCp} has similar parameters as that of {\tt relxill} with the addition of log$N$ (cm$^{-3}$) to vary the density of the accretion disc with an upper limit of $20$ and electron temperature ($kT_{e}$) of the comptonizing plasma instead of $E_{cut}$. We kept some of the model parameters fixed during the fit as mentioned for the {\tt relxill} model. This reflection model also improved the fit significantly to $\chi^2/dof=1242/1267$ ($\Delta\chi^2=-253$ for the $5$ additional parameters). The corresponding spectrum for the model {\tt constant*TBabs*(bbody+relxillCp)} and the residuals are shown in the right panel of Figure~\ref{Fig4}. The best-fit parameter values using both reflection models are consistent and fully compatible with each other. All the best-fit parameter values are listed in Table~\ref{parameters1}. Previously, \citet{2019MNRAS.483..767D} employed this model successfully to describe the spectral properties of the known AMSP SAX~J1808.4-3658.\\

Most interestingly, the reflection component implies a large disc truncation in both models: the $1\:\sigma$ upper limit on $R_{in}\simeq 6\:R_{ISCO}$ for the {\tt relxill } model and  $R_{in}\simeq 9\:R_{ISCO}$ for the {\tt relxillCp} model. Both models also yield a consistent, moderate inclination estimate of $37\degr_{-7}^{+8}$ for the {\tt relxill} model and $40\degr_{-8}^{+12}$ for the {\tt relxillCp} model. We estimated a high disc ionization of log$\xi=4.12_{-0.15}^{+0.10}$. The value is consistent with the typical range observed in NS LMXBs (log $\xi\sim 3-4$). Both reflection fits appear to favor a high Fe abundance. The $1\:\sigma$ lower limit on Fe abundance lies $\simeq 1.58$. This could be indicative of a high density of the accretion disc. The {\tt relxillCp} model predicted an upper limit of the disc density parameter of $10^{18}$ cm$^{-3}$. The reflection fraction $f_{refl}$ parameter represents the ratio of illuminating flux to that which is reflected for {\tt relxill} and {\tt relxillCp}. The value of reflection fraction $f_{refl}$ is close to unity, as expected, since the majority of emitted photons would interact with the disk with relatively fewer emitted towards infinity. \\

\begin{figure*}
\includegraphics[scale=0.40, angle=0]{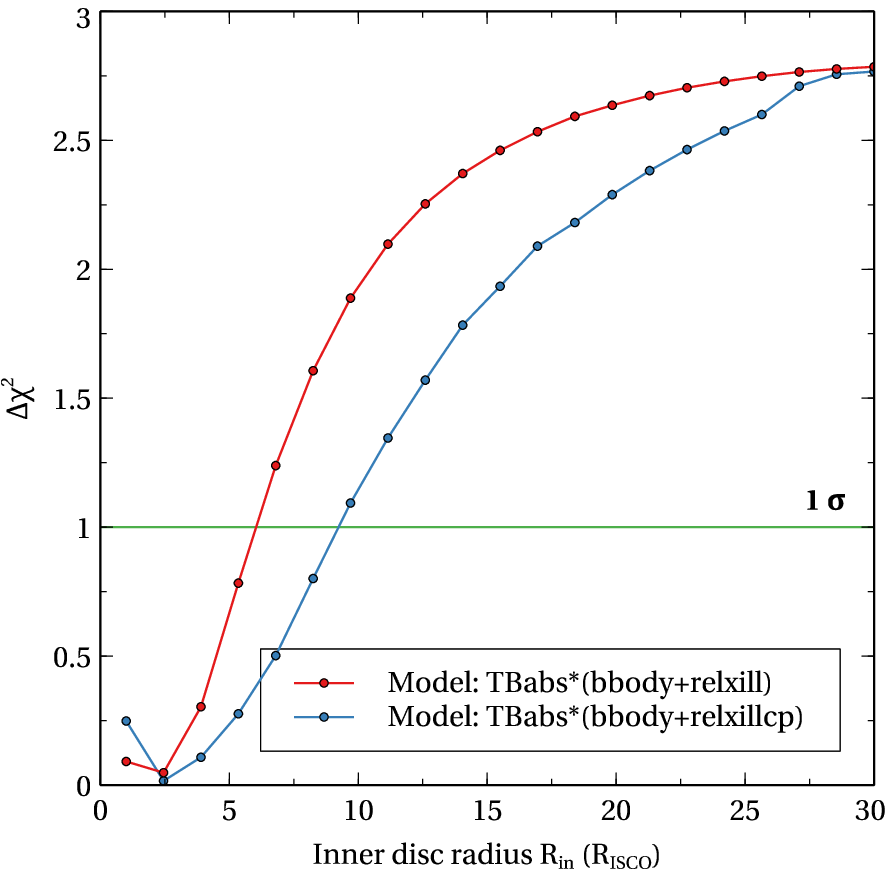}\hspace{2cm}
\includegraphics[scale=0.40, angle=0]{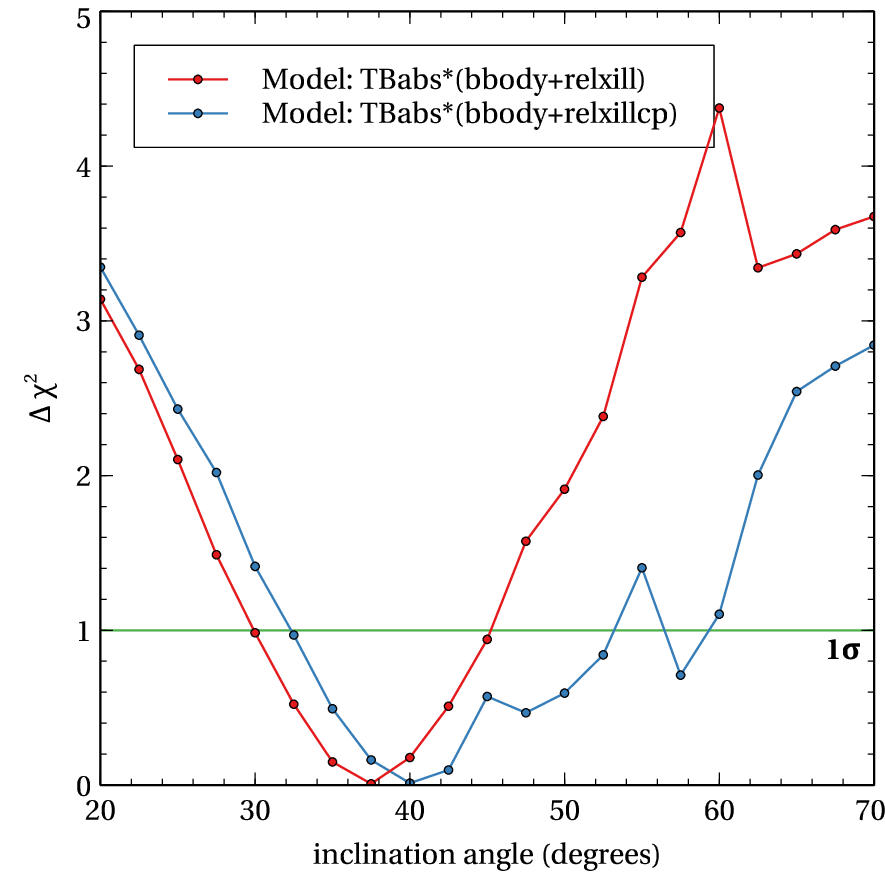}
\caption{The plots show the change in the goodness of fit for the inner disc radius ($R_{in}$) and disc inclination angle ($i$). The left panel shows the variation of $\Delta\chi^{2}(=\chi^{2}-\chi_{min}^{2})$ as a function of $R_{in}$ (varied between $1$ to $30\:R_{ISCO}$) obtained from two different model combination. The right panel shows the variation of $\Delta\chi^{2}(=\chi^{2}-\chi_{min}^{2})$ as a function of $i$ obtained from different models. We varied the disc inclination angle between $20$ degrees and $70$ degrees.} 
\label{Fig6}
\end{figure*}

Further, we have tested the goodness of the fit for the inner disk radius, $R_{in}$, and disc inclination angle, $i$. We used the command {\tt steppar} in {\tt xspec} to search the best-fit for $R_{in}$ and $i$ for both best-fit models {\tt const*TBabs*(bbody+relxill)} and {\tt const*TBabs*(bbody+relxillCp)}. Left panel of Figure~\ref{Fig6} shows the variation of $\Delta\chi^{2}(=\chi^{2}-\chi_{min}^{2})$ as a function of $R_{in}$ for both models and right panel of Figure~\ref{Fig6} shows the variation of $\Delta\chi^{2}$ with $i$ for both models. \\

\subsection{Type I X-Ray Burst}
During \nustar{} observation of IGR~J17498-2921, two type-I thermonuclear bursts have been observed in the $3-79$ keV lightcurve. We have performed an analysis of individual bursts and showed them in Figure~\ref{Fig8} and Figure~\ref{Fig9}. Both bursts have sharp rises and exponential falls and typically last 20 seconds. The rising part has hardly been detected because of its rapidness ($<$ 1 sec). Therefore, we divided the burst into three segments: the peak of the burst (part-1; 5 sec around the peak flux), the exponential fall (part-2; next 5 sec after the peak), and the tail of the burst (part-3; last 10 sec of the burst as it gradually flattens). Three segments of the bursts are shown by stars, circles, and triangles in the top left panel of Figure \ref{Fig8} and \ref{Fig9}, respectively. We have created GTI files, extracted spectra and performed analyses for three segments separately. \\

We found that burst spectra extracted from FPMA and FPMB during three segments can jointly be fitted by a single blackbody radiation ({\tt bbodyrad} in {\tt xspec}) in the presence of neutral absorption ({\tt TBabs} in {\tt xspec}). The blackbody nature of the burst has also been observed in several past studies. A constant is used to normalize the response between FPMA and FPMB. The absorption column density is kept fixed at $0.81 \times$ 10$^{22}$ cm$^{-2}$, which is the line-of-sight column density. Best-fit spectral parameters for individual segments are provided in Table~\ref{parameter2} for both bursts. The top right, bottom left, and bottom right panels of Figure~\ref{Fig8} and \ref{Fig9} show best-fit unfolded spectra along with residual during peak flux, exponential fall, and tail of both bursts, respectively. \\
 
From Table~\ref{parameter2}, we may note that the blackbody temperature is highest during the peak of the burst and falls off significantly as the burst count falls towards the tail. Similarly, the blackbody emitting radius (a proxy for the blackbody normalization) is highest at the peak and becomes smaller as the flux decreases. A similar trend in temperature and radius has also been observed in other type-I X-ray bursts (\citealt{2008ApJS..179..360G, 2021ASSL..461..209G, 2022ApJ...935L..32B, 2024ApJ...968L...7N}).

 \begin{table*}
   \centering
\caption{Fit results: Best-fitting spectral parameters of the \nustar{} observation of the source IGR~J17498-2921.\\
          Model 1:  {\tt const*TBabs*(bbody+relxill)} and Model 2:{\tt const*TBabs*(bbody+relxillCp)}} 
\begin{tabular}{|p{2.8cm}|p{5.0cm}|p{3.2cm}|p{3.2cm}}
    \hline
    Component     & Parameter (unit) & Model 1 & Model 2 \\
    \hline
    {\scshape Constant} & FPMB (wrt FPMA) & $0.98\pm 0.003$ & $0.98\pm 0.002$ \\
    {\scshape tbabs}    & $N_{H}$($\times 10^{22}\;\text{cm}^{-2}$) & $0.81 $(f)  & $0.81 $(f)   \\
    {\scshape bbody} & $kT_{bb} (\kev)$ & $1.61_{-0.10}^{+0.21}$    & $1.66_{-0.08}^{+0.13}$ \\
    & Norm ($\times 10^{-3}$)& $1.01_{-0.12}^{+0.20}$   &  $1.11_{-0.12}^{+0.10}$ \\  
    {\scshape relxill/relxillCp} & $i$ (degrees) & $37_{-7}^{+8}$ & $40_{-8}^{+12}$ \\
    & $R_{in}$ ($\times R_{ISCO}$) & $\lesssim 6$ & $\lesssim 9$ \\
    & $\rm{log}\:\xi$ (erg cm s$^{-1}$) &  $4.12_{-0.15}^{+0.10}$ & $4.28_{-0.48}^{+0.07}$ \\
    & $\Gamma$  & $1.68\pm 0.02 $ & $1.76\pm 0.03 $ \\
    & $A_{Fe}$ ($\times \;\text{solar})$   & $\gtrsim 1.58$  & $\gtrsim 2.14$\\
    & $E_{cut}(\kev)$ &  $105_{-20}^{+12}$ & -- \\
    & log$N$ (cm$^{-3}$) &  -- & $\lesssim 18$\\
    & $kT_{e}  (\kev)$ & -- & $19_{-1}^{+2}$ \\
    & $f_{refl}$   & $4.21_{-3.06}^{\dagger}$ & $1.13_{-0.40}^{+1.32}$ \\
    & norm ($\times 10^{-4}$)   &  $3.80_{-1.21}^{+1.98}$ & $9.29_{-2.3}^{+1.9}$\\
    {\scshape cflux} & $F_{bbody}^{*}$ ($\times 10^{-9}$ ergs/s/cm$^2$) & $0.08\pm 0.02$ & $0.09\pm 0.02$\\
    & $F_{relxill/relxillCp}^{*}$ ($\times 10^{-9}$ ergs/s/cm$^2$) & $1.04\pm 0.01$ & $1.03\pm 0.01$\\
    & $F_{total}^{*}$ ($\times 10^{-9}$ ergs/s/cm$^2$) & $1.12\pm 0.01$ & $1.12\pm 0.01$\\
   \hline 
    & $\chi^{2}/dof$ & $1244/1268$  & $1242/1267$ \\
    \hline
  \end{tabular}\label{parameters1} \\
{\bf Note:} The outer radius of the {\tt relxill/relxillCp} spectral component was fixed to $1000\;R_{g}$. We fixed emissivity index $q1=q2=3$. The spin parameter ($a$) was fixed to $0.19$ as the spin frequency of the NS is $401$ Hz. $^{*}$All the unabsorbed fluxes are calculated in the energy band $3-79 \kev{}$ using the {\tt cflux} model component in {\tt XSPEC}. $^{\dagger}$ upper bound error calculation is not well constrained.\\

\end{table*}

\begin{figure*}
    \centering
\includegraphics[angle=-90,scale=0.32]{fig8a.ps}
\includegraphics[angle=-90,scale=0.32]{fig8b.ps}
\includegraphics[angle=-90,scale=0.32]{fig8c.ps}
\includegraphics[angle=-90,scale=0.32]{fig8d.ps}
\caption{Top Left: Lightcurve profile of the first type I burst is shown in 3-20 keV as observed by \nustar{}/FPMA (top) and \nustar{}/FPMB (bottom). The peak of the profile, the falling part, and the tail are shown in stars, circles, and triangles. FPMA and FPMB joint best-fit spectra analyzed during the peak profile, falling part and the tail of burst-1 are shown in the top right, bottom left, and bottom right panels, respectively, along with the residuals.}
\label{Fig8}
\end{figure*}

\begin{figure*}
\centering
\includegraphics[angle=-90,scale=0.32]{fig9a.ps}
\includegraphics[angle=-90,scale=0.32]{fig9b.ps}
\includegraphics[angle=-90,scale=0.32]{fig9c.ps}
\includegraphics[angle=-90,scale=0.32]{fig9d.ps}
\caption{Top Left: Lightcurve profile of the second type I burst is shown in 3-20 keV as observed by \nustar{}/FPMA (top) and \nustar{}/FPMB (bottom). The peak of the profile, the falling part, and the tail are shown in stars, circles, and triangles. FPMA and FPMB joint best-fit spectra were analyzed during the peak profile, the falling part and the tail of burst-2 are shown in the top right, bottom left, and bottom right panels, respectively, along with the residuals.}
\label{Fig9}
\end{figure*}

\begin{table}
\centering
\caption{Burst spectral analysis: best-fit spectral parameters during three segments (Part 1-3) of both bursts are provided. kT$_{bbody}$ is the blackbody model temperature in keV, and BBodyNorm is the blackbody normalization. }
\label{parameter2}
\begin{tabular}{llccc} 
\hline
Burst & Part & kT$_{bbody}$ & BBody & $\chi^2/dof$\\
Number  & number & (keV) & Norm & \\
\hline
& Part-1 & 2.19$^{+0.05}_{-0.04}$ & 177$^{+14}_{-13}$ & 62/63 \\
BURST-1 & Part-2 & 1.78$^{+0.05}_{-0.05}$ & 106$^{+12}_{-11}$ & 36/34 \\
& Part-3 & 1.38$^{+0.04}_{-0.04}$ & 94$^{+13}_{-11}$ & 23/27 \\
\hline
& Part-1 & 2.14$^{+0.06}_{-0.05}$ & 204$^{+17}_{-18}$ & 41/46 \\
BURST-2 & Part-2 & 2.03$^{+0.05}_{-0.05}$ & 93$^{+13}_{-11}$ & 23/30 \\
& Part-3 & 1.52$^{+0.06}_{-0.06}$ & 76$^{+12}_{-11}$ & 22/20 \\
\hline
\end{tabular}
\end{table}

\section{Discussion}
We present the results of the spectral analysis of the AMSP IGR~J17498-2921 using the \nustar{} observation during its decaying phase of 2023 outbursts. During this observation, the source is detected with an average count rate of $\sim 22\ctssec$ . The unabsorbed $3-79\kev{}$ flux is estimated at $1.1\times 10^{-9}$ erg~s$^{-1}$ cm$^{-2}$, consistent to that of \nicer{} flux in the $0.5-10$\kev{} energy band observed on 19th April 2023 \citep{2023ATel15998....1S}. However, The unabsorbed bolometric X-ray flux during this observation in the energy band $0.1-100\kev{}$ is $1.32\times 10^{-9}$ erg~s$^{-1}$ cm$^{-2}$. This implies an unabsorbed bolometric luminosity of $7.8\times 10^{36}$ erg~s$^{-1}$, assuming a distance of $7.6$ kpc. This value corresponds to $\sim 2\%$ of the Eddington luminosity ($L_{Edd}$) which is $\sim 3.8\times 10^{38}$ erg~s$^{-1}$ for a canonical $1.4\:M_{\odot}$ NS \citep{2003A&A...399..663K}. The X-ray luminosity during outburst remains below $10\%$ the $L_{Edd}$ in the vast majority of the AMSPs. From the \inte{} data, maximum flux observed during the 2023 outburst was around $4.7\times 10^{-9}$ erg~s$^{-1}$ cm$^{-2}$ in the energy band $28-84\kev{}$ \citep{2023ATel15996....1G}. So, the unabsorbed source flux decays roughly four times within a short timescales (a span of around ten days). The peak flux observed during the 2011 outburst in the 2-30 keV energy band was $\sim 1.1\times 10^{-9}$ erg~s$^{-1}$ cm$^{-2}$, smaller than the 2023 peak outburst flux. \\

We found that the continuum X-ray emission is well described using an absorbed blackbody ($kT_{bb}=1.55 \pm 0.03\kev{}$) and a power-law ($\Gamma= 1.70\pm 0.02$) with a cut-off energy at $ 55\pm 4 \kev{}$. The combination of absorbed blackbody and a physically motivated Comptonization model {\tt nthcomp} also described the continuum fairly well with electron temperature $kT_{e}=16\pm 2\kev{}$, photon index $\Gamma= 1.79\pm 0.02$, and the seed photon temperature $kT_{seed}\lesssim 0.31\kev{}$ likely originating from the disk. Thus, to describe the continuum emission of IGR~J17498-2921, we need a single blackbody component of temperature $kT_{bb}\simeq 1.6\kev{}$, likely arising from the NS surface and a Comptonization spectrum which can be associated with a hot corona scattering of photons from a cold disc of temperature $kT_{seed}\lesssim 0.31\kev{}$, whose direct emission is likely too faint for detection. The blackbody parameters ($kT_{bb}\sim 1.6$\kev{}, $R_{bb}\sim 1.1-1.2$ km) suggest that this continuum might originate from a small part of the NS surface. From continuum modeling, we found that the spectrum is dominated by a hard Comptonization component, comprising of $\simeq 88\%$ of the total $3 - 79$ \kev{} unabsorbed flux, suggesting a behavior similar to the other AMSPs (\citealt{2020A&A...641A..37K, 2021A&A...649A..76L, 2022MNRAS.516L..76S, 2023MNRAS.519.3811S, 2023ApJ...958..177L}). The electron temperature of the corona, i.e., $\sim 16$\kev{}, is also comparable with the temperature found for other AMSPs in hard spectral states. The spectrum exhibited the presence of broad Fe-K emission line in $5-9\kev{}$, a Compton hump $12-30\kev{}$, indicating the disc reflection of the coronal emission (i.e., hard X-ray photons) subject to the strong relativistic distortion \citep{1989MNRAS.238..729F}\\

The observed features of the continuum modeling motivate us to employ the disc reflection models. We found that the $3-60\kev{}$ source energy spectrum is adequately fitted using a model combination consisting of an absorbed single-temperature blackbody model ({\tt bbody}) and a reflection model {\tt relxill} or {\tt relxillCp}. Both reflection model infers a large disc truncation radius: the $1\:\sigma$ upper limit on $R_{in}\simeq 6\:R_{ISCO}$ for the {\tt relxill } model and  $R_{in}\simeq 9\:R_{ISCO}$ for the {\tt relxillCp} model. Both models yield a consistent, moderate inclination estimate of $37\degr_{-7}^{+8}$ for the {\tt relxill} model and $40\degr_{-8}^{+12}$ for the {\tt relxillCp} model, which is consistent with the fact that no dips or eclipses have been observed in the light curve of this source. Similar disc inclination angle has also been observed for the other AMSPs such as IGR~J17511-3057, HETE~J1900-2455, and SAX~J1748.9-2021 (\citealt{2010MNRAS.407.2575P, 2013MNRAS.429.3411P, 2016MNRAS.457.2988P}). The reflection fit also revealed that the accretion disc is highly ionized with log$\xi\sim 4.0$, resulting in a strong reflection continuum. The value is consistent with the typical range observed in both black hole and NS LMXBs (log($\xi$) $\sim 3-4$). Moreover, both reflection models produce consistent ionization measurements. The iron abundance is also greater than the solar composition, the lower limit of $A_{Fe}\simeq 1.58$. High value of log$\xi$ and $A_{Fe}$ could indicate the high density of the disc \citep{2018ApJ...855....3L}. From the reflection model {\tt relxillCp}, we measured an upper limit of the density in the disc surface of log$(n_{e}/\rm{cm}^{-3})\lesssim 18$.\\

The upper limit of the inner disc radius $R_{in}$ obtained from the reflection model {\tt relxill} and {\tt relxillCp} are $\sim 6 \:R_{ISCO}$ and $\sim 9 \:R_{ISCO}$ (where $R_{ISCO}=5.4\:R_{g} \sim 11.4$ km), respectively, which corresponds to $\sim 68$ km and $\sim 102$ km, respectively. The lower limit of $R_{in}$ can be obtained from the highest observed orbital frequency at the inner edge of the accretion disc. The source IGR~J17498-2921 exhibits a coherent signal at a frequency of $401$ Hz. Following \citet{1998ApJ...508..791M}, we estimated an lower limit of $R_{in}$ to first order in $j$ using the following equation 
\begin{equation}
R_{in}\geq 19.5 (\nu_{orb}/1000\: \text{Hz})^{-1}(1+0.2j) \text{km}
\end{equation} 
Here $j=cJ/GM^{2}$ is the dimensionless angular momentum (spin parameter) of the NS. The above equation estimates a lower limit of $R_{in}\sim 51$ km, corresponding to $\sim 4.5\: R_{ISCO}$. This constraint depends upon the detection of the coherent signal or QPO frequency from the NS system. Despite this fact, it allows us to put a constraint on $R_{in}$ which is $\simeq(4.5-6)\:R_{ISCO}$ from the {\tt relxill} model and $\simeq (4.5-9)\:R_{ISCO}$ from the {\tt relxillCp} model. This estimation is consistent with most of the other AMSPs for which a spectral analysis has been performed, and a broad iron line has been detected in moderately high resolution spectra, such as IGR~J17511-3057 \citep{2010MNRAS.407.2575P}, IGR~J17480-2446 \citep{2011ApJ...731L...7M}, HETE~J1900.1-2455 \citep{2013MNRAS.429.3411P}, SAX~J1748.9-2021 \citep{2016MNRAS.457.2988P}, IGR~J17062-6143 \citep{2021ApJ...912..120B}, and Swift~J1749.4-2807 \citep{2022MNRAS.515.3838M}.\\

We further used the best-fit spectral parameters to compute physical properties like mass accretion rate ($\dot{m}$), the maximum radius of the boundary layer ($R_{BL,max}$), and magnetic field strength ($B$) of the NS in the system. We first estimated the mass accretion rate per unit area, using Equation (2) of \citet{2008ApJS..179..360G}
\begin{equation}
\begin{split}
\dot{m}=&\:6.7\times 10^{3}\left(\frac{F_{p}\:c_\text{bol}}{10^{-9} \text{erg}\: \text{cm}^{-2}\: \text{s}^{-1}}\right) \left(\frac{d}{10 \:\text{kpc}}\right)^{2} \left(\frac{M_\text{NS}}{1.4 M_{\odot}}\right)^{-1}\\
 &\times\left(\frac{1+z}{1.31}\right) \left(\frac{R_\text{NS}}{10\:\text{km}}\right)^{-1} \text{g}\: \text{cm}^{-2}\: \text{s}^{-1}.
 \end{split} 
\end{equation}
The above equation yields a mass accretion rate of $1.2\times 10^{-9}\;M_{\odot}\;\text{y}^{-1}$ at a persistent flux $F_{p}=1.1\times 10^{-9}$ erg~s$^{-1}$ cm$^{-2}$, assuming the bolometric correction $c_{bol} \sim 1.38$ \citep{2008ApJS..179..360G}. Here we assume $1+z=1.31$ (where $z$ is the surface redshift) for an NS with mass ($M_{NS}$) 1.4 $M_{\odot}$ and radius ($R_{NS}$) $10$ km. We used Equation (2) of \citet{2001ApJ...547..355P} to estimate the maximum radial extension of the boundary layer from the NS surface based on the mass accretion rate. We found the maximum value of the boundary layer to extend to $R_{BL}\sim 5.52\;R_{g}\: (1.02 \:R_{ISCO})$, assuming $M_{NS}=1.4\:M_{\odot}$ and $R_{NS}=10$ km. The extent of the boundary layer region is small but consistent with the disc position. However, the actual value may be larger than this if we account for the spin and the viscous effects in this layer. \\

For X-ray pulsars, the magnetic field of an NS can potentially truncate the inner disc and re-direct plasma along the magnetic field lines. The upper limit of $R_{in}$ measured from the reflection fit can be used to estimate an upper limit of the magnetic field strength of the NS. Equation (1) of \citet{2009ApJ...694L..21C} gives the following expression for the magnetic dipole moment,
\begin{equation}
\begin{split}
\mu=&3.5\times 10^{23}k_{A}^{-7/4} x^{7/4} \left(\frac{M}{1.4 M_{\odot}}\right)^{2}\\
 &\times\left(\frac{f_{ang}}{\eta}\frac{F_{bol}}{10^{-9} \text{erg}\: \text{cm}^{-2}\: \text{s}^{-1}}\right)^{1/2}
 \frac{D}{3.5\: \text{kpc}} \text{G}\; \text{cm}^{3}.
\end{split} 
\end{equation}
We assumed a geometrical coefficient $k_{A}=1$, an anisotropy correction factor $f_{ang}=1$, and the accretion efficiency in the Schwarzschild metric $\eta=0.2$ (as reported in \citealt{2009ApJ...694L..21C} and \citealt{2000AstL...26..699S}). We used $0.1 - 100$\kev{} flux as the bolometric flux ($F_{bol}$) of $\sim 1.32\times 10^{-9}$ erg~s$^{-1}$ cm$^{-2}$. $R_{in}$ is modified as $R_{in}=x\:GM/c^{2}$ by introducing a scale factor $x$ (\citet{2009ApJ...694L..21C}). Using the upper limit of $R_{in}$ ($\lesssim 33\:R_{g}$) from the {\tt relxill} fit, we obtained $\mu \leq 0.9\times 10^{27}$ G cm$^{3}$. This corresponds to an upper limit of the magnetic field strength of $B\lesssim 1.8\times 10^{9}$ G at the magnetic poles (assuming an NS mass of $1.4\:M_{\odot}$, a radius of $10$ km, and a distance of $7.6$ kpc). This is perfectly consistent with the previously estimated magnetic field strength of IGR~J17498-2921 by \citet{2015MNRAS.452.3994M}. \\

Furthermore, two type-I X-ray bursts have been observed during the \nustar{} observation. This indicates that the source is still accreting during the \nustar{} observation, even when the disk is truncated at a significant distance. The behavior is quite similar with the other NS LMXBs and AMSPs (\citealt{2010MNRAS.407.2575P, 2016MNRAS.457.2988P, 2016ApJ...819L..29K, 2017MNRAS.466L..98V}). We performed the spectral analysis of the bursts after dividing it into three segments: the peak of the burst, the exponential fall, and the tail of the burst. We modeled the burst emission with an absorbed blackbody. The burst temperature peaks at the start of the burst and falls off significantly as the burst count falls toward the tail. The burst decay is well described by a cooling thermal blackbody component, which supports the thermonuclear origin for the X-ray burst (like other AMSPs SRGA~J144459.2-604207 and  MAXI~J1816–195, \citealt{2024ApJ...968L...7N, 2022ApJ...935L..32B}). The inferred blackbody radius during the peak of these X-ray bursts is estimated to be $\sim 10\pm 2.7$ km and $\sim 11\pm 3$ km for burst1 and burst2, respectively. For the source, IGR~J17498-2921, the way the temperature and radius change during the bursts are commonly observed for other AMSPs as well, but some AMSPs show the deviation of burst spectra from the standard blackbody model (\citealt{2022ApJ...936L..21C, 2024ApJ...966L...3J}). To further investigate the source properties, one needs to examine the evolution of the type I X-ray burst morphology. For those, intense X-ray monitoring observations are required. \\
 
\subsection*{Postscript}
Two similar papers appeared online after we archived the present manuscript and submitted it to this journal on 12th August 2024. One paper by \citet{2024arXiv240806895I} appeared on 13th August 2024, a day after the submission of our paper. The other one by \citet{2024arXiv240812786L} appeared on 23rd August 2024. To the best of our knowledge, these papers have yet to be published. Still, in the following two paragraphs, we discuss the similarities and dissimilarities between their findings and ours.\\

Both analyzed the spectrum of the source IGR~J17498-2921 using the same \nustar{} observation, in combination with \nicer{}. However, \citet{2024arXiv240812786L} used the \inte{} data along with \nustar{} and \nicer{} for spectral analysis. Both works identified the source to be in the hard spectral state. \citet{2024arXiv240806895I} found that the broad-band spectral emission ($1-79$\kev{}) observed quasi-simultaneously by \nicer{} and \nustar{} is well described by an absorbed Comptonized emission plus a disc reflection component. They observed that the Comptonized continuum was well described by a photon index of $\Gamma \sim 1.95$, an electron temperature of $kT_{e} \sim 17$\kev{}, a seed photons temperature of $kT_{seed} \sim 0.60$, which are perfectly consistent with our findings. They adopted a self-consistent reflection model {\tt rfxconv} with the relativistic blurring kernel {\tt rdblur} to model the reflection component. From the reflection modeling, they obtained an upper limit of the inner accretion disc radius $R_{in}\sim 38\;R_{g}$ with a system inclination of $\gtrsim 60\degr$. The estimated inner accretion disc radius is consistent with our estimation. However, we estimated a lower system inclination ($\sim 40\degr$), which is consistent with \citet{2012A&A...545A..26F} and in agreement with the absence of dips or eclipses in the light curve. \citet{2024arXiv240806895I} further obtained a higher ionization parameter of the disc of log$\xi\sim 3.3$, keeping $A_{Fe}$ fixed at the solar value. \\

\citet{2024arXiv240812786L} analyzed the joint quasi-simultaneous \nicer{}, \nustar{} and \inte{} spectra in the energy range of $1-150$\kev{}. They found that the spectra are well described by a self-consistent reflection model, {\tt relxillCp}, with a Gaussian line of instrumental origin, modified by interstellar absorption. They reported that the Comptonized emission from the hot corona is characterized by a photon index $\Gamma$ of $\sim 1.8$ and an electron temperature $kT_{e}$ of $\sim 39$\kev{}, implying a hard spectral state. They obtained a low inclination angle $i\sim 34\degr$. Their broad-band spectral fitting further showed the properties of the accretion disk as a strong ionization, log($\xi$/erg cm s$^{-1}$) $\sim 4.5$, over-solar abundance, $A_{Fe}\sim 7.7$, and high density, log($n_{e}$/cm$^{-3}$ ) $\sim 19.5$. All these estimations, including inclination angle, are consistent with our findings from the reflection modeling with \nustar{} observation. However, they estimated an upper limit of the inner accretion disc radius $R_{in}\sim 2.1\;R_{ISCO}$, suggesting that the accretion disc is located rather close to the NS surface. They further noticed that the addition of an extra blackbody component with a temperature of $\sim 1.6\kev{}$ changed the inner disc radius $R_{in}$ to $4.5^{+10.7}_{-2.1} R_{ISCO}$ and the electron temperature $kT_{e}$ to $30\pm 8\kev{}$. Moreover, \citet{2024arXiv240812786L} constrained the magnetic field of IGR~J17498-2921 in the range of $(0.9 - 2.4) \times 10^{8}$ G, which is compatible with our estimation. We note some discrepancies in the estimation of some parameters between our and the findings of \citet{2024arXiv240806895I} and \citet{2024arXiv240812786L}. However, the values are still comparable with each other within the uncertainty. There are some discrepancy in the measurement of $R_{in}$ as different reflection models have been used to estimate the same, producing varying estimates (see also Table 3 of \citet{2024arXiv240806895I}). It demonstrates the complexity of fitting the reflection components. Moreover, our estimation of $R_{in}$ is based on the \nustar{} spectra solely in the energy range $3-60\kev{}$, while other studies derived $R_{in}$ from \nustar{} spectra, in combination with \nicer{} and sometimes \inte{}. It further reflects the importance of multi-instrument observations and advanced spectral modeling for precisely measuring the accretion disc radius.

\section{Data availability}
This research has made use of data obtained from the HEASARC, provided by NASA's Goddard Space Flight Center. The observational data sets with Obs. IDs $90901317002$ (\nustar{}) dated April 23, 2023 and $6203770103$ (\nicer{}) dated April 23, 2023 are in public domain put by NASA at their website https://heasarc.gsfc.nasa.gov.  
 
\section{Acknowledgements}
 We thank the anonymous referee for comments that improved the description of the results. This research has made use of data and/or software provided by the High Energy Astrophysics Science Archive Research Centre (HEASARC). This research also has made use of the \nustar{} data analysis software ({\tt NuSTARDAS}) jointly developed by the ASI Space Science Data Center (SSDC, Italy) and the California Institute of Technology (Caltech, USA). ASM and BR would like to thank Inter-University Centre for Astronomy and Astrophysics (IUCAA) for their facilities extended to him under their Visiting Associate Programme.

\def\apj{ApJ}
\def\apjl{ApJl}
\def\pasp{PASP} \def\mnras{MNRAS} \def\aap{A\&A} \def\physerp{PhR} \def\apjs{ApJS} \def\pasa{PASA}
\def\pasj{PASJ} \def\nat{Nature} \def\memsai{MmSAI} \def\araa{ARAA} \def\iaucirc{IAUC} \def\aj{AJ} \def\aaps{A\&AS}
\bibliographystyle{mn2e}
\bibliography{aditya}

\appendix
\section{\nicer{} observation}
\begin{figure*}
\centering
\includegraphics[scale=0.50, angle=0]{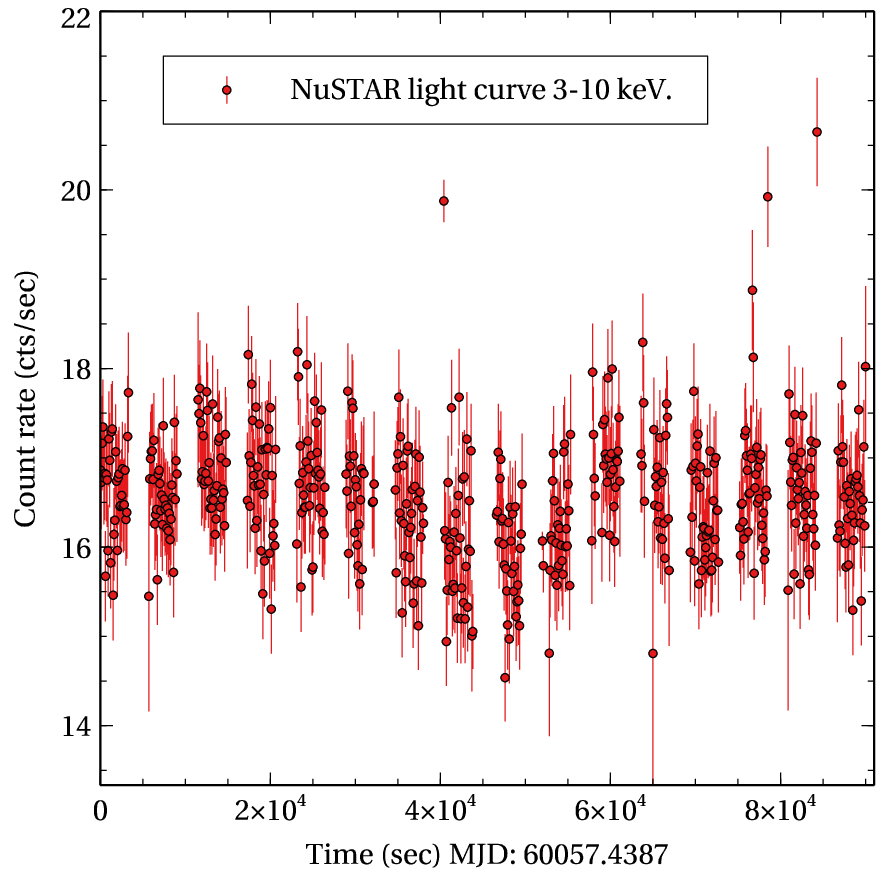}\hspace{2cm}
\includegraphics[scale=0.50, angle=0]{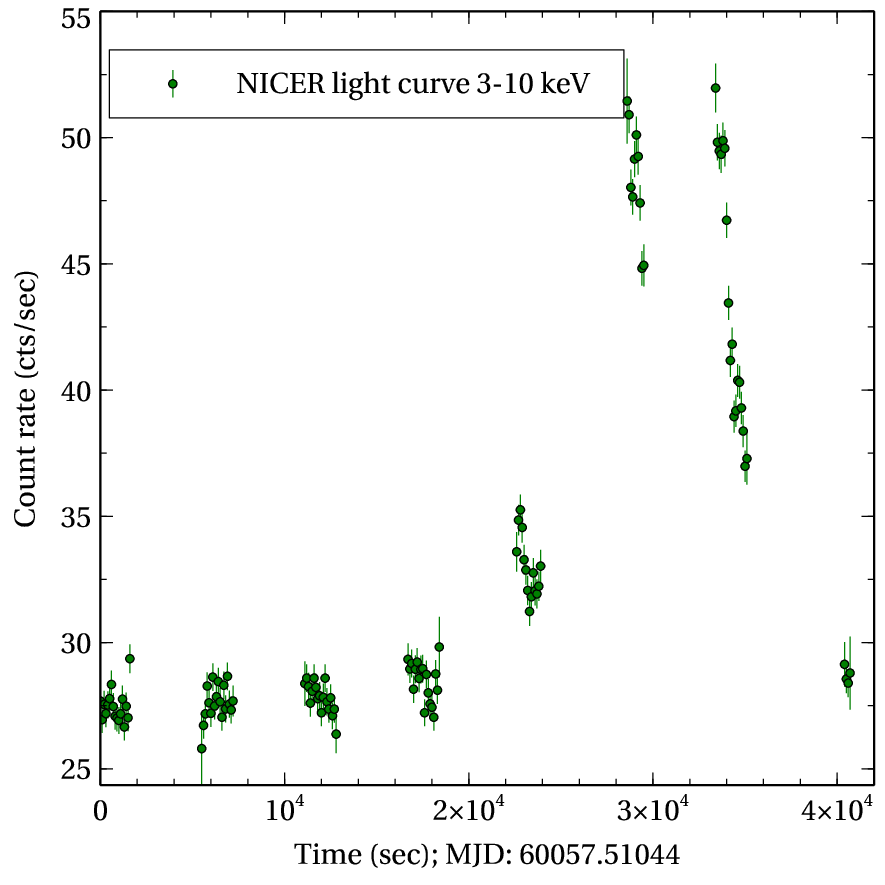}
\caption{The left panel shows the \nustar{} light curve in the $3-10$ \kev{} energy band with 100s binning. The right panel shows a similar light curve with \nicer{} detector. A certain rise in the count rate ($\sim 50$ cts/s) is observed in the \nicer{} lightcurve, which was absent during the \nustar{} observation.} 
\label{Fig10}
\end{figure*}

After the reported outburst of IGR~J17498-2921 on April 13-15, 2023, \nicer{} conducted frequent observations. The source was observed by \nicer{} on April 23, 2023, which was carried out simultaneously with \nustar{} observation. The \nicer{} observation started after $\sim 6.3$ ks of the \nustar{} observation. Here, we report on this public \nicer{} observation with ObsID $6203770103$ \citep{2016SPIE.9905E..1HG}. We analyzed this observation using the {\tt nicerdas v12} pipeline in {\tt HEAsoft v6.33} and {\tt CALDB xti20240206}. We used {\tt nicerl2} task to produce standard, calibrated, cleaned event files for this \nicer{} observation. The light curves of this source are generated using the task {\tt nicer-lc}. The $3-10$\kev{} \nicer{} light curve of $100$ s binning is shown in the right panel of Figure~\ref{Fig10}. For comparison, we have also shown the \nustar{} light curve in a similar energy band in the left panel of Figure~\ref{Fig10}. We found a mismatch between simultaneous \nustar{} and \nicer{} lightcurves. Both lightcurves are background-subtracted following standard procedures. We did not observe any hump-like structure (a certain rise in the count rate $\sim 50$ cts/s after $\sim 20$ ks from the beginning of the observation) in the \nustar{} lightcurve that was present in the \nicer{} lightcurve, although they are simultaneous and even in similar energy bands. That's why we did not include the \nicer{} spectrum for further analysis.

\end{document}